\documentclass[12pt]{article}
\usepackage{graphicx,amssymb,bbold}
\usepackage{amsmath,amsfonts,epsfig}

\newcommand{\be}{\begin{equation}}
\newcommand{\ee}{\end{equation}}
\newcommand{\bea}{\begin{eqnarray}}
\newcommand{\eea}{\end{eqnarray}}
\newcommand{\beas}{\begin{eqnarray*}}
\newcommand{\eeas}{\end{eqnarray*}}

\newcommand{\nn}{\nonumber}

\begin{document}
\begin{titlepage}

\vspace*{-24mm}
\rightline{YITP-13-48}
\vspace{4mm}

\begin{center}

{\bf {\Large Black Hole Formation in}} \\[3pt]
\vspace{1mm}
{\bf {\Large Fuzzy Sphere Collapse}}

\vspace{8mm}

\renewcommand\thefootnote{\mbox{$\fnsymbol{footnote}$}}
Norihiro Iizuka${}^{1}$\footnote{iizuka@yukawa.kyoto-u.ac.jp},
Daniel Kabat${}^{2}$\footnote{daniel.kabat@lehman.cuny.edu},
Shubho Roy${}^{3}$\footnote{sroy@cts.iisc.ernet.in} and
Debajyoti Sarkar${}^{2,4}$\footnote{dsarkar@gc.cuny.edu}

\vspace{4mm}

${}^1${\small \sl Yukawa Institute for Theoretical Physics} \\
{\small \sl Kyoto University, Kyoto 606-8502, JAPAN}

${}^2${\small \sl Department of Physics and Astronomy} \\
{\small \sl Lehman College, City University of New York, Bronx NY 10468, USA}

${}^3${\small \sl Center for High Energy Physics} \\
{\small \sl Indian Institute of Science, Bangalore 560012, INDIA}

${}^4${\small \sl Graduate School and University Center} \\
{\small \sl City University of New York, New York NY 10036, USA}

\end{center}

\vspace{4mm}

\noindent
We study the collapse of a fuzzy sphere, that is a spherical membrane built out of D0-branes,
in the BFSS model.  At weak coupling, as the sphere shrinks, open strings are produced.
If the initial radius is large then open string production is not important and
the sphere behaves classically.  At intermediate initial radius the back-reaction from
open string production is important but the fuzzy sphere retains its identity.  At small initial
radius the sphere collapses to form a black hole.  The crossover between the later two regimes
is smooth and occurs at the correspondence point of Horowitz and Polchinski.

\end{titlepage}
\setcounter{footnote}{0}
\renewcommand\thefootnote{\mbox{\arabic{footnote}}}

\section{Introduction\label{sect:intro}}

Recently in \cite{Iizuka:2013yla} we studied bound state formation in D-brane collisions, including the
possible formation of a black hole.  We considered collisions between clusters of D-branes, as well as
a configuration in which D-branes were arranged in a spherical shell with velocities directed toward the
center.  At weak coupling a bound state forms via a process of open string production.  At strong
coupling, where the system has a dual supergravity description \cite{Itzhaki:1998dd}, the collision results in formation
of a black hole.  We found that the crossover between these two mechanisms for bound state formation
is smooth.  It occurs at an intermediate value of the coupling, in accord with the correspondence principle
introduced by Horowitz and Polchinski \cite{Horowitz:1996nw}.

The purpose of the present paper is to study a more interesting initial configuration, namely a fuzzy sphere or
spherical membrane built out of 0-branes.  Starting from rest, a fuzzy sphere will shrink under its own tension.
Classically the sphere shrinks to zero size and re-expands.  But taking quantum effects into account, as the
sphere shrinks open string production can occur at weak coupling, while black hole formation can occur at
strong coupling.  Our objective is to study these processes in more detail and show that they are smoothly
connected at the correspondence point.

An outline of this paper is as follows.  In \S \ref{sect:fuzzy} we review the description of fuzzy spheres and study
the spectrum of fluctuations about a fuzzy sphere.  In \S \ref{sect:perturbative} we study the collapse of a fuzzy
sphere at weak coupling as open strings are produced.  In \S \ref{sect:BHformation} we argue that there is a smooth
match to the process of black hole formation at strong coupling.  In \S \ref{sect:parametric} we study the perturbative
evolution of the sphere in more detail, including back-reaction from open string production.  In \S \ref{sect:more} we
provide further evidence for a smooth crossover at the correspondence point.

There is a large literature on fuzzy geometry in various matrix models, for a review see \cite{Steinacker:2011ix}.  
For studies of thermalization and black hole formation in these models see for example \cite{Berenstein:2010bi,Asplund:2011qj,Riggins:2012qt,Asplund:2012tg}.

\section{Fuzzy spheres\label{sect:fuzzy}}

To describe an ordinary sphere embedded in ${\mathbb R}^d$, we begin by introducing three
Cartesian coordinates $x_A = (x,y,z)$ on a unit $S^2$, subject to the constraint
\[
\sum_A  x_A^2 = 1
\]
The embedding coordinates in ${\mathbb R}^d$, which we denote $X^i$ for $i = 1,\ldots,d$, can then
be expanded in powers of the $x_A$'s.
\be
X^i = \sum_{\ell = 0}^\infty c^i_{A_1 \cdots A_\ell} \, x_{A_1} \cdots x_{A_\ell}
\ee
The coefficients $c^i_{A_1 \cdots A_\ell}$ are symmetric and traceless on their lower indices.  They transform in the spin-$\ell$ representation of $SU(2)$.
After the traces are removed, the product $x_{A_1} \cdots x_{A_\ell}$ provides a Cartesian basis for the spin-$\ell$ spherical harmonics \cite{Sakurai}.

To make the sphere fuzzy or non-commutative we use the dictionary \cite{Madore:1991bw,Balachandran:2005ew,Aschieri:2006uw}
\be
x_A \leftrightarrow {2 \over N} J_A
\ee
where the matrices $J_A$ are generators of $SU(2)$ in the $N$-dimensional representation (i.e.\ with spin $j = {N - 1 \over 2}$).  They obey
\be
[J_A,J_B] = i \epsilon_{ABC} J_C \qquad\quad \sum_A J_A^2 = {N^2 - 1 \over 4} {\mathbb 1}
\ee
The embedding coordinates become Hermitian matrices, with the expansion
\be
X^i = \sum_{\ell = 0}^{N-1} c^i_{A_1 \cdots A_\ell} \left({2 \over N}\right)^\ell J_{A_1} \cdots J_{A_\ell}
\ee
Note that the expansion terminates at $\ell = N-1$, since beyond this point one no longer gets
independent matrices.  To make this plausible, note that summing the dimensions of the appropriate $SU(2)$ representations
accounts for the $N^2$ parameters in a Hermitian matrix.
\be
\sum_{\ell = 0}^{N-1} (2 \ell + 1) = N^2
\ee
In fact there is a stronger result: the matrices vanish identically for $\ell \geq N$.  To see this it's convenient to work in a basis of raising and lowering operators $J_\pm = J_x \pm i J_y$ with metric $ds^2 = dx^+ dx^- + dz^2$.  Note that $(J_+)^\ell$ is traceless and symmetric on its lower indices -- it's the highest weight state in the spin-$\ell$ representation -- and with $N$-dimensional generators it vanishes
identically for $\ell \geq N$, $(J_+)^\ell = 0$ for $\ell \geq N$.  Then by applying lowering operators a general symmetrized traceless product must vanish
for $\ell \geq N$.

This construction of a fuzzy sphere finds a natural home in the BFSS model \cite{Banks:1996vh}, or the quantum
mechanics of $N$ D0-branes, where the bosonic part of the action is\footnote{Conventions: the fields $X^i$ have units of energy.  They are related to 0-brane positions
by $X = ({\rm position}) / 2 \pi \alpha'$.  The Yang-Mills coupling is
$g^2_{\rm YM} = {g_s \over (2\pi)^2 \ell_s^3}$.\label{conventions}}
\be
S = {1 \over g^2_{\rm YM}} \int dt \, {\rm Tr} \left( {1 \over 2} (\partial_0 X^i)^2
+ {1 \over 4} [X^i, X^j]^2 \right)
\ee
We've fixed the gauge $A_0 = 0$, so the equation of motion
\be
\ddot{X}^i + [[X^i,X^j],X^j] = 0
\ee
must be supplemented with the Gauss constraint
\be
\label{Gauss}
[\partial_0 X^i,X^i] = 0
\ee

At the classical level a simple configuration is a spherical membrane of initial radius $U_0$, described
by setting \cite{Kabat:1997im,Rey:1997iq}
\bea
\label{ClassicalSphere}
&& X^A(t) = U(t) {2 \over N} J^A \qquad A = 1,2,3 \\[2pt]
\nonumber
&& X^I = 0 \hspace{29mm} I = 4,\cdots,9
\eea
The Gauss constraint is trivially satisfied since $[J^A,J^A] = 0$, while the equation of motion
reduces to
\be
\label{BackgroundEOM}
\ddot{U} = - {8 \over N^2} U^3
\ee
Solving this with the initial conditions $U(0) = U_0$, $\dot{U}(0) = 0$ one finds that the sphere collapses after a time
\be
\label{ClassicalTimescale}
\tau = {N \Gamma(1/4)^2 \over \sqrt{128 \pi} \, U_0}
\ee
This construction of a spherical membrane is based on the pioneering work of de Wit et al.\ \cite{deWit:1988ig}.  The collapsing sphere solution was first
described by Collins and Tucker \cite{Collins:1976eg}.

In the quantum theory we'll be interested in fluctuations about this solution, so we set\footnote{The results for the spectrum in the remainder of this section
have also been obtained by Harold Steinacker and Jochen Zahn \cite{Harold}.}
\bea
&& X^A(t) = U(t) {2 \over N} J_A + x^A(t) \\
\nonumber
&& X^I(t) = x^I(t)
\eea
At linearized order the Gauss constraint (\ref{Gauss}) reduces to
\be
\label{LinearizedGauss}
\dot{U} [J^A,x^A] = U [J^A,\dot{x}^A]
\ee
This constraint removes roughly $N^2$ degrees of freedom from the $3N^2$ degrees of freedom contained in $x^A$.\footnote{More precisely it removes $N^2 - 1$ degrees of freedom: the trace of a commutator vanishes, so the trace of the Gauss constraint is trivially satisfied.}  However to linearized order it puts no constraint on $x^I$.

The linearized equation of motion for $x^I$ is
\be
\label{LinearizedxIeom}
\ddot{x}^I + {4 \over N^2} U^2 [J^A,[J^A,x^I]] = 0
\ee
To solve this we expand the field in fuzzy spherical harmonics,
\be
\label{xIexpansion}
x^I = \sum_{\ell = 0}^{N-1} c^I_{A_1 \cdots A_\ell} \left({2 \over N}\right)^\ell J_{A_1} \cdots J_{A_\ell}
\ee
With the $SU(2)$ algebra $[J_A,J_B] = i \epsilon_{ABC} J_C$ and the identity $\epsilon_{ABC} \epsilon_{ADE} = \delta_{BD} \delta_{CE} - \delta_{BE} \delta_{CD}$ one can show that (assuming the indices $A_1 \cdots A_\ell$ are contracted with a symmetric traceless tensor)
\be
\label{FuzzyCasimir}
[J_A,[J_A,J_{A_1} \cdots J_{A_\ell}]] = \ell (\ell + 1) J_{A_1} \cdots J_{A_\ell}
\ee
In other words, fuzzy spherical harmonics are angular momentum eigenstates, with the expected eigenvalue of the total angular momentum.
The linearized equation of motion (\ref{LinearizedxIeom}) then reduces to
\be
\label{FluctuationEquation}
\ddot{c}^I_{A_1 \cdots A_\ell} + {4 \ell (\ell + 1) \over N^2} U^2 c^I_{A_1 \cdots A_\ell} = 0
\ee
This determines the spectrum of fluctuations in the transverse dimensions $I = 4,\ldots,9$.  In each of these dimensions there are fluctuations with
$\ell = 0,\ldots,N-1$.  A fluctuation with angular momentum $\ell$ is $(2\ell+1)$-fold degenerate and has frequency
\be
\label{omega_ell}
\omega_\ell = {2 \over N} \sqrt{\ell (\ell + 1)} \, U
\ee
The spectrum of fluctuations in the dimensions $A = 1,2,3$ is studied in appendix \ref{appendix:xA}.  Here we just summarize the
results.  Decomposing $x^A$ into $SU(2)$ representations we find that
there are $s$-type fluctuations with spin $\ell + 1$ for $\ell = 0,\ldots,N-1$.  
These fluctuations are $(2\ell + 3)$-fold degenerate and have frequency
\be
\label{somega_ell}
\omega_\ell = {2 \over N} \sqrt{\ell (\ell - 1)} \, U
\ee
There are also $u$-type fluctuations with spin $\ell - 1$ for $\ell = 1,\ldots,N-1$.  These fluctuations are $(2\ell - 1)$-fold
degenerate and have frequencies
\be
\label{uomega_ell}
\omega_\ell = {2 \over N} \sqrt{(\ell + 1) (\ell + 2)} \, U
\ee
In the rest of this paper the distinction between these various types of frequencies will not matter, and from now on we will ignore the
differences between the formulas (\ref{omega_ell}), (\ref{somega_ell}), (\ref{uomega_ell}).  When we write explicit formulas we
will make use of the transverse frequencies (\ref{omega_ell}).

\begin{table}
\begin{center}
\begin{tabular}{|c|c|c|c|c|}
\hline
name & labels & spin & degeneracy & frequency \\
\hline
transverse & $\displaystyle \rule{0pt}{12pt} I = 4,\ldots,9 \atop \displaystyle \rule{0pt}{10pt} \ell = 0,\ldots,N-1$ & $\ell$ & $2 \ell + 1$ & ${2 \over N} \sqrt{\ell(\ell+1)} \, U$ \\[6pt]
\hline
$s$-type & $\rule{0pt}{14pt} \ell = 0,\ldots,N-1$ & $\ell + 1$ & $2 \ell + 3$ & ${2 \over N} \sqrt{\ell(\ell-1)} \, U$ \\[5pt]
\hline
$u$-type & $\rule{0pt}{14pt} \ell = 1,\ldots,N-1$ & $\ell - 1$ & $2 \ell - 1$ & ${2 \over N} \sqrt{(\ell+1)(\ell+2)} \, U$ \\[4pt]
\hline
\end{tabular}
\end{center}
\caption{Spectrum of fluctuations about a fuzzy sphere of radius $U$.}
\end{table}

\section{Perturbative sphere collapse\label{sect:perturbative}}

Assuming the 0-brane quantum mechanics is weakly coupled, let's study the collapse of a fuzzy sphere in a little more detail.  The
conserved total energy of the quantum mechanics is
\be
E_{\rm YM} = {1 \over g^2_{\rm YM}} {\rm Tr} \left( {1 \over 2} (\partial_0 X^i)^2 - {1 \over 4} [X^i, X^j]^2\right)
\ee
which at large $N$ for the classical solution (\ref{ClassicalSphere}) reduces to
\be
\label{EYM}
E_{\rm YM} \approx {1 \over g^2_{\rm YM}} \left({N \over 2} \dot{U}^2 + {2 \over N} U^4\right)
\ee
So the radial velocity $\dot{U}$ is related to the initial radius of the sphere $U_0$ by
\be
\label{velocity}
\dot{U}^2 \approx {4 \over N^2} \big( U_0^4 - U^4 \big)
\ee

Classically a fuzzy sphere remains spherical as it collapses, but quantum mechanically other modes will get excited.  This happens
when the adiabatic approximation breaks down.  For a mode with frequency $\omega_\ell$, the adiabatic approximation fails when
\be
{\dot{\omega}_\ell \over \omega_\ell^2} \gtrsim 1
\ee
Given the frequencies (\ref{omega_ell}), adiabaticity breaks down when
\be
{N \dot{U} \over U^2 \sqrt{\ell(\ell + 1)}} \gtrsim 1
\ee
which using (\ref{velocity}) can be rewritten as
\be
U \lesssim {U_0 \over \big(\ell(\ell+1) + 1\big)^{1/4}}
\ee
So a large fuzzy sphere evolves adiabatically.  As the sphere shrinks modes with more and more angular momentum become excited.
The mode with the largest angular momentum, $\ell_{\rm max} \sim N$, gets excited when the fuzzy sphere reaches the inner radius for open string production
\be
\label{Uinner}
U_{\rm inner} \sim U_0 / \sqrt{N}
\ee
At this point the adiabatic approximation has completely broken down, and all $N^2$ degrees of freedom in the matrices have become excited,
or equivalently all possible open strings have been produced.
The subsequent evolution of the sphere will be studied in section \ref{sect:parametric}.

\section{Black hole formation and the correspondence point\label{sect:BHformation}}

At large $N$ and strong coupling the 0-brane quantum mechanics has a dual description in terms of IIA supergravity \cite{Itzhaki:1998dd}.
Introducing the 't Hooft coupling $\lambda = g^2_{\rm YM} N$ and a radial coordinate with units of energy $U = r / \alpha'$, the 0-brane
quantum mechanics is weakly coupled when $U > \lambda^{1/3}$ and has a dual supergravity description when $U < \lambda^{1/3}$.\footnote{The
radial coordinate $U$ introduced here differs by a factor $2\pi$ from the radius of the sphere introduced in (\ref{ClassicalSphere}): see
footnote \ref{conventions}.  We will ignore
this difference from now on.}

In the supergravity regime one would expect a fuzzy sphere to collapse and form a black hole with $N$ units of 0-brane charge.  The Schwarzschild
radius of such a black hole is \cite{Itzhaki:1998dd}
\be
\label{US}
U_S \sim \big(g_{\rm YM}^4 E\big)^{1/7}
\ee
Here $E$ is the energy above extremality, identified with the Hamiltonian of the quantum mechanics.  Given (\ref{EYM}), the Schwarzschild radius is related
to the initial radius of the sphere by
\be
U_S \sim \left({g_{\rm YM}^2 U_0^4 \over N}\right)^{1/7}
\ee
Of course this discussion only makes sense if the black hole fits in the region where supergravity is valid.  This requires $U_S < \lambda^{1/3}$ or equivalently
$E < N^2 \lambda^{1/3}$, which means
\be
U_0 < N^{1/2} \lambda^{1/3}
\ee

The perturbative description of fuzzy sphere collapse worked out in section \ref{sect:perturbative}, on the other hand, is only valid if the quantum mechanics
is weakly coupled.  We followed the evolution of the sphere perturbatively down to the radius $U_{\rm inner}$ given in (\ref{Uinner}), at which point all
$N^2$ degrees of freedom have gotten excited.  This perturbative description is only valid if $U_{\rm inner} > \lambda^{1/3}$, or equivalently
\be
U_0 > N^{1/2} \lambda^{1/3}
\ee

We now see that there is a smooth crossover between the perturbative description of fuzzy sphere collapse and the non-perturbative process of black
hole formation.  The crossover occurs when the initial radius and total energy are
\bea
&& U_0 \sim N^{1/2} \lambda^{1/3} \\
\nonumber
&& E \sim N^2 \lambda^{1/3}
\eea
At the crossover point the Schwarzschild radius and inner radius for open string production agree,
\be
U_S \sim U_{\rm inner} \sim \lambda^{1/3}
\ee
For a black hole of this size the curvature at the horizon is of order string scale.
\be
\alpha' R \sim (U^3 / \lambda)^{1/2} \sim 1
\ee
So this crossover is an example of the correspondence principle of Horowitz and Polchinski at work \cite{Horowitz:1996nw}.

\section{Back-reaction and parametric resonance\label{sect:parametric}}

In \S \ref{sect:perturbative} we followed the evolution of a weakly-coupled fuzzy sphere down to the radius
$U_{\rm inner} \sim U_0 / \sqrt{N}$.  At this radius adiabaticity has broken down for all of the fluctuation modes, so
${\cal O}(N^2)$ open strings have been produced.  In this section we study the subsequent evolution of the sphere, still assuming
weak coupling, but taking into account back reaction from open string production.  We'll show that a
parametric resonance is present in the weakly-coupled field theory which exponentially amplifies the number of open strings
present.

To study the back-reaction from open string production, we begin by estimating the total energy in open strings.
Suppose that as the sphere collapses roughly one open string is produced in each of the fluctuation modes (\ref{omega_ell}).  This is
justified in appendix \ref{appendix:open}.  Then once the sphere has crossed the radius $U_{\rm inner}$, the total energy in open strings is
\bea
\nonumber
E_{\rm open} & \sim & \sum_{\ell = 0}^{N-1} (2 \ell + 1) \, \omega_\ell \\
\nonumber
& = & \sum_{\ell = 0}^{N-1} (2 \ell + 1) {2 \over N} \sqrt{\ell(\ell + 1)} \, U \\
\label{Eopen}
& \sim & N^2 U
\eea
For this description of the collapse process to make sense, we should check that back-reaction from open
string production can be neglected down to the radius $U_{\rm inner}$.  To do this we compare the energy in open strings (\ref{Eopen})
to the total energy of the sphere (\ref{EYM}).  At the radius $U_{\rm inner}$ we have $E_{\rm open} \sim N^2 U_{\rm inner}$, while
the total energy $E_{\rm YM} \sim U_0^4 / \lambda \sim N^2 U_{\rm inner}^4 / \lambda$, so
\be
{E_{\rm open} \over E_{\rm YM}} \sim {\lambda \over U_{\rm inner}^3}
\ee
Indeed, provided the field theory remains weakly coupled down to the radius $U_{\rm inner}$, we have $U_{\rm inner} > \lambda^{1/3}$
(or equivalently $U_0 > N^{1/2} \lambda^{1/3}$) and back reaction can be neglected during the initial collapse of the sphere.

Even though back-reaction can be neglected during the initial collapse of the sphere, it is not necessarily negligible when the sphere
subsequently re-expands.  To decide this issue we compare the potential energy in open strings (\ref{Eopen}), $E_{\rm open} \sim N^2 U$,
to the classical potential energy
of a fuzzy sphere, which from (\ref{EYM}) is given by $E_{\rm classical} = {2 \over \lambda} U^4$.  Thus
\be
{E_{\rm open} \over E_{\rm classical}} \sim {N^2 \lambda \over U^3}
\ee
The linear potential from open strings
dominates at small radius, while the classical $U^4$ potential dominates at large radius.  The two energies are comparable when $U \sim N^{2/3} \lambda^{1/3}$.

We can now identify three qualitatively different behaviors, depending on the initial radius of the sphere. 
See appendix  \ref{appendix:moreonU(t)} for a more detailed analysis.

\goodbreak
\noindent
\underline{{\em large initial radius,} $U_0 > N^{2/3} \lambda^{1/3}$}

\noindent
In this case the classical potential energy of the sphere is dominant near the turning point, which is located at $U \approx U_0$.  The classical evolution of the sphere
described in \S \ref{sect:fuzzy} is a good approximation to the true behavior.  In particular the sphere collapses to zero size on the timescale
$\tau \sim N / U_0$ given in (\ref{ClassicalTimescale}).

\noindent
\underline{{\em intermediate initial radius,} $N^{1/2} \lambda^{1/3} < U_0 < N^{2/3} \lambda^{1/3}$}

\noindent
In this case the field theory remains weakly coupled down to the radius $U_{\rm inner}$, but when the sphere subsequently re-expands it's the linear potential
arising from open string production which is dominant near the turning point.  The classical $U^4$ potential can be neglected, and
overall energy conservation reads (in place of (\ref{EYM}))
\be
{N \over 2 g^2_{\rm YM}} \dot{U}^2 + c N^2 U = {2 U_0^4 \over \lambda}
\ee
Here  $c$ is an ${\cal O}(1)$ constant reflecting the number of open strings present in each mode.  In this linear potential the turning point
is located at $U = 2 U_0^4 / c N^2 \lambda$, which fortunately is in the weakly coupled regime of the field theory.  After reaching
the turning point, the sphere re-collapses to zero size in a time
\be
\tau = {2 U_0^2 \over c N \lambda}
\ee

\goodbreak
\noindent
\underline{{\em small initial radius,} $U_0 < N^{1/2} \lambda^{1/3}$}

\noindent
In this case the sphere enters the regime where supergravity is valid and falls within its own Schwarzschild radius to form a black hole.

\bigskip

We can now describe the subsequent evolution of the sphere in a little more detail. 
At weak coupling the sphere pulsates with a frequency
\be
\label{Omega}
\Omega \sim 1/\tau \sim \left\lbrace
\begin{array}{ll}
U_0 / N & \hbox{\rm large initial radius} \\[3pt]
N \lambda / U_0^2 & \hbox{\rm intermediate initial radius}
\end{array}
\right.
\ee
One can approximate this as an oscillating classical background $U(t) = \widetilde{U}_0 \sin \Omega t$, where the back-reacted amplitude of oscillation
\be
\label{U0tilde}
\widetilde{U}_0 \sim \left\lbrace
\begin{array}{ll}
U_0 & \hbox{\rm large initial radius} \\[3pt]
U_0^4 / N^2 \lambda & \hbox{\rm intermediate initial radius}
\end{array}
\right.
\ee
Plugging this oscillating background into the fluctuation equation (\ref{FluctuationEquation}) 
for the transverse fluctuations, one finds that small fluctuations are governed
by the Mathieu equation.  
As in \cite{Iizuka:2013yla}, this means there is a parametric resonance which makes the number of open strings
grow exponentially with time, on a timescale set by the 
period of oscillation $\tau$.\footnote{Similarly, for $s$-type and $u$-type fluctuations, 
we obtain Mathieu equations with $\omega_l$ given by 
(\ref{somega_ell}) and (\ref{uomega_ell}). 
However the derivation of (\ref{somega_ell}) and (\ref{uomega_ell}) in 
appendix \ref{appendix:xA} is under the 
adiabatic approximation, $\dot{U} \to 0$. Therefore 
we expect the Mathieu equations for $s$-type and $u$-type fluctuations are modified 
once the adiabatic approximation breaks down and parametric resonance occurs.}

\section{More on the correspondence point\label{sect:more}}

The collapse of a fuzzy sphere appears qualitatively different depending on whether the initial radius is large, intermediate or small.
In this section we study the transitions between these different regimes, and argue that they are in fact smoothly connected.

One can smoothly continue from large to intermediate
initial radius in the formulas (\ref{Omega}), (\ref{U0tilde}) for the frequency and amplitude of oscillation, since the expressions agree at the large-to-intermediate crossover point $U_0 \sim N^{2/3} \lambda^{1/3}$.  In a way this is not surprising.  
At large and intermediate initial radius open string production takes place while the field theory is still weakly coupled.  As the initial radius is decreased open string
production becomes more important.  The resulting linear potential smoothly takes over from the classical $U^4$ potential, and this is responsible for modifying
the frequency and amplitude of oscillation.

Now let's see if we can continue from intermediate to small initial radius.  This intermediate-to-small crossover occurs when $U_0 \sim N^{1/2} \lambda^{1/3}$,
which corresponds to a total energy $E \sim U_0^4 / \lambda \sim N^2 \lambda^{1/3}$.  This amounts to working at the correspondence point of Horowitz and
Polchinski \cite{Horowitz:1996nw}, since the Schwarzschild radius of the resulting black hole
\be
\label{UScorrespondence}
U_S \sim (g_{\rm YM}^4 E)^{1/7} \sim \lambda^{1/3}
\ee
which means the curvature at the horizon is of order string scale.
\be
\alpha' R \sim (U_S^3 / \lambda)^{1/2} \sim 1
\ee
In other words, the resulting black hole just fits in the region where supergravity is valid \cite{Itzhaki:1998dd}.

There are various quantities we can compare at the Horowitz-Polchinski correspondence point which suggest that the crossover is smooth.

\goodbreak
\noindent
\underline{\em classical size}

\noindent
In the weakly-coupled field theory the classical background is a pulsating sphere with a maximum size given in (\ref{U0tilde}).  Evaluating this at
$U_0 = N^{1/2} \lambda^{1/3}$ we find that the back-reacted amplitude of oscillation is set by the 't Hooft scale,
$\widetilde{U}_0 \sim \lambda^{1/3}$.  This matches the Schwarzschild radius (\ref{UScorrespondence}) of a black hole at the correspondence point,
$U_S \sim \lambda^{1/3}$.

\goodbreak
\noindent
\underline{\em size of quantum fluctuations}

\noindent
For the classical background (\ref{ClassicalSphere}), the size of the sphere can be measured by
\be
{1 \over N} {\rm Tr} \left(X^A X^A\right) = U^2 \left(1 + {\cal O}(1/N^2)\right)
\ee
Let's compare this to the spread in the 0-brane positions due to quantum fluctuations, measured by
\be
(\Delta X)^2 \equiv {1 \over N} \langle {\rm Tr} \left(X^I X^I\right) \rangle \qquad I = 4,\ldots,9
\ee
To evaluate this, recall that for a harmonic oscillator
\be
\langle n \vert \hat{x}^2 \vert n \rangle = {\hbar \over m \omega} \left(n + {1 \over 2}\right)
\ee
We can adapt this to the problem at hand by identifying $\hbar/m$ with $g^2_{\rm YM}$.  Then assuming small fluctuations
and using the frequencies (\ref{omega_ell}) we have
\bea
\nonumber
(\Delta X)^2 & = & \sum_I {1 \over N} \sum_{\ell = 1}^{N-1} \sum_{m = -\ell}^\ell {g^2_{\rm YM} \over \omega_\ell} \left(n_{\ell m}^I + {1 \over 2}\right) \\
\nonumber
& \sim & {1 \over N} \sum_{\ell = 1}^{N-1} (2 \ell + 1) {g^2_{\rm YM} N \over \sqrt{\ell(\ell+1)} \, U} \\
& \sim & {\lambda \over U}
\eea
In the first line we suppressed the $\ell = 0$ modes which describe center of mass position.  In the second line we dropped
the sum on $I$ and took
the quantum numbers $n^I_{\ell m} \sim {\cal O}(1)$, appropriate to having one open string per mode.  To compare the size
of these quantum fluctuations to the size of the classical background, we set $U = \widetilde{U}_0$ and consider the ratio
\be
{(\Delta X)^2 \over (\widetilde{U}_0)^2} \sim {\lambda \over (\widetilde{U}_0)^3}
\ee
Provided the maximum size of the sphere is larger than the 't Hooft scale, $\widetilde{U}_0 > \lambda^{1/3}$ or equivalently
$U_0 > N^{1/2} \lambda^{1/3}$, then the quantum fluctuations in the 0-brane positions are small compared to the radius of the
sphere.  This shows that at large and intermediate initial radius a classical fuzzy sphere provides a good description of the quantum state.\footnote{Although as we saw in \S \ref{sect:parametric}, for intermediate initial radius one must take back-reaction into account to find the correct frequency and amplitude for the classical background.}  It also shows
that as we go to the Horowitz-Polchinski correspondence point, $\widetilde{U}_0 = \lambda^{1/3}$, the classical background
merges into the quantum fluctuations.  This fits with a general expectation in gravity-gauge duality, that at strong
coupling the D-brane positions have quantum fluctuations which are comparable in size to the region in which supergravity
is valid \cite{Susskind:1998vk,Polchinski:1999br}.

\goodbreak
\noindent
\underline{\em thermalization time}

\noindent
On the weakly-coupled side we identified a parametric resonance which leads to open string production on a timescale set by the frequency
(\ref{Omega}).  Evaluating this at
$U_0 = N^{1/2} \lambda^{1/3}$ we find that the frequency of oscillation is set by the 't Hooft scale,
$\Omega \sim \lambda^{1/3}$.

What does this correspond to on the supergravity side?
The black hole has a spectrum of quasinormal frequencies which govern the approach to equilibrium.
The quasinormal frequencies are set by the Hawking temperature \cite{Horowitz:1999jd,Iizuka:2003ad,Maeda:2005cr}, namely
$T \sim {1 \over \sqrt{\lambda}} U_S^{5/2}$,
which at the correspondence point is of order the 't Hooft scale, $T \sim \lambda^{1/3}$.
Thus at the correspondence point the timescale for parametric resonance agrees with the relaxation time of the black hole.  This suggests that the
weak-coupling process of open string production via parametric resonance smoothly matches on to the strong-coupling process of black hole formation.

\goodbreak
\noindent
\underline{\em entropy production}

\noindent
At weak coupling, during the initial collapse of a fuzzy sphere, we saw that ${\cal O}(N^2)$ open strings are produced.  These strings have
an entropy $S_{\rm string}
\sim N^2$.
On the other hand, on the supergravity side, the equilibrium entropy of the black hole is \cite{Itzhaki:1998dd}
\be
S_{\rm bh} \sim N^2 U_S^{9/2} / \lambda^{3/2}
\ee
Evaluating this at the correspondence point $U_S \sim \lambda^{1/3}$ we see that $S_{\rm bh} \sim N^2$.  So at the correspondence point the entropy produced during
the initial collapse of a fuzzy sphere is close to the equilibrium entropy of the black hole.  This suggests that very little additional evolution -- perhaps just
a few e-foldings of parametric resonance -- is required for the system to reach equilibrium.

\bigskip
\goodbreak
\centerline{\bf Acknowledgements}
\noindent
We are grateful to Harold Steinacker for valuable discussions.  NI was supported in part by JSPS KAKENHI Grant Number 25800143.  DK and DS were
supported in part by U.S.\ National Science Foundation grants PHY-0855582 and PHY-1125915 and by grants from
PSC-CUNY.  The research of SR is supported in part by Govt.\ of India Department of Science and Technology's research grant under scheme DSTO/1100 (ACAQFT).

\appendix
\section{Fluctuations in the $X^A$ dimensions\label{appendix:xA}}

In this appendix we study the spectrum of fluctuations in the directions $A = 1,2,3$.
We need to solve the linearized Gauss constraint
\be
\dot{U} [J^A,x^A] = U [J^A,\dot{x}^A]
\ee
along with the linearized equation of motion
\be
\label{xAeom}
\ddot{x}^A + {4 \over N^2} U^2[[x^A,J^B],J^B] + {4 \over N^2} U^2 [[J^A,x^B],J^B] + {4 \over N^2} U^2 [[J^A,J^B],x^B] = 0
\ee
These expressions can be simplified somewhat.  In the adiabatic approximation we study the spectrum of fluctuations treating $U$ as constant.  Then the fluctuation
modes can be taken to have definite frequency, $x^A \sim e^{-i \omega t}$, so the Gauss constraint amounts to the requirement that
\be
\label{Gauss2}
[J^A,x^A] = 0
\ee
Also we can simplify the equation of motion using
\beas
[[J^A,x^B],J^B] &=& - [[x^B,J^B],J^A] - [[J^B,J^A],x^B] \qquad \hbox{\rm (Jacobi identity)} \\
&=& [[J^A,J^B],x^B] \hspace{4cm} \hbox{\rm (Gauss constraint)}
\eeas
This reduces the equation of motion to
\be
\label{xAeom2}
\ddot{x}^A + {4 \over N^2} U^2[[x^A,J^B],J^B] + {8 \over N^2} U^2 [[J^A,J^B],x^B] = 0
\ee

To go further we expand the fluctuations in fuzzy vector spherical harmonics.  These are constructed as follows.
Expanding in a complete set of matrices we can set\footnote{To save writing we're adopting a different normalization convention
in expanding $x^A$,
without the factor $\left(2 \over N\right)^\ell$ present in (\ref{xIexpansion}).}
\be
x^A = \sum_{\ell = 0}^{N-1} x_{AA_1 \cdots A_\ell}  J_{A_1} \cdots J_{A_\ell}
\ee
The tensor $x_{AA_1 \cdots A_\ell}$ is symmetric and traceless on the indices $A_1 \cdots A_\ell$, so taking all indices
into account it transforms as a $(\hbox{\rm spin 1}) \otimes (\hbox{\rm spin $\ell$})$ product representation of $SU(2)$.
Decomposing this product, the irreducible pieces correspond to tensors
$s,t,u$ that have spin $\ell + 1$, $\ell$, $\ell - 1$ respectively.  These tensors can be constructed explicitly.\footnote{A hat denotes a missing
index.  There's an
overall normalization in these formulas which we leave unspecified.}
\bea
\nonumber
s_{A_0 A_1 \cdots A_\ell} &\sim& \left(x_{A_0 A_1 \cdots A_\ell} + \hbox{\rm cyclic permutations of $A_0 \cdots A_\ell$}\right) \\
&& - {2 \over 2 \ell + 1} \sum_{\scriptstyle i,j = 0 \atop \scriptstyle i < j}^\ell \delta_{A_i A_j} x_{BBA_0 \cdots \widehat{A}_i \cdots
\widehat{A}_j \cdots A_\ell} \\
\label{projector}
t_{A_1 \cdots A_\ell} &\sim& \sum_{i = 1}^\ell \epsilon_{A_iAB} x_{ABA_1 \cdots \widehat{A}_i \cdots A_\ell} \\[2pt]
u_{A_2 \cdots A_\ell} &\sim& \delta_{AB} x_{ABA_2 \cdots A_\ell}
\eea
The tensors $s,t,u$ are constructed to be symmetric and traceless on all indices, so that they correspond to the appropriate
irreducible $SU(2)$ representations.

This decomposition helps in understanding the Gauss constraint (\ref{Gauss2}), since
\bea
\nonumber
&& \hspace{-1.5cm} [J^A,x^A] = x_{AA_1 \cdots A_\ell} [J^A, J_{A_1} \cdots J_{A_\ell}] \\
\nonumber
&&= i \left(\epsilon_{A_1AB} x_{ABA_2 \cdots A_\ell} + \epsilon_{A_2AB} x_{AA_1BA_3 \cdots A_\ell} + \cdots
+ \epsilon_{A_\ell AB} x_{AA_1 \cdots A_{\ell - 1} B}\right) J_{A_1} \cdots J_{A_\ell} \\
\nonumber
&&\sim i t_{A_1 \cdots A_\ell} J_{A_1} \cdots J_{A_\ell}
\eea
Thus the Gauss constraint requires that we set the spin-$\ell$ irreducible piece to zero, $t_{A_1 \cdots A_\ell} = 0$.

Now let's study the equation of motion (\ref{xAeom2}).  Using (\ref{FuzzyCasimir}) in the middle term, and evaluating the commutators in
the last term, the equation of motion becomes
\bea
\label{xAeom3}
&& \ddot{x}_{AA_1 \cdots A_\ell} J_{A_1} \cdots J_{A_\ell} + {4 \over N^2} U^2 \ell(\ell+1) x_{AA_1 \cdots A_\ell} J_{A_1} \cdots J_{A_\ell} \\
\nonumber
&& + {8 \over N^2} U^2 x_{BBA_2 \cdots A_\ell} \left(J_A J_{A_2} \cdots J_{A_\ell} + J_{A_2} J_A J_{A_3} \cdots J_{A_\ell} + \cdots\right) \\[2pt]
\nonumber
&& - {8 \over N^2} U^2 x_{BAA_2 \cdots A_\ell} \left(J_B J_{A_2} \cdots J_{A_\ell} + J_{A_2} J_B J_{A_3} \cdots J_{A_\ell} + \cdots\right) \\[2pt]
\nonumber
&& = 0
\eea
(there are $\ell$ terms in the second and third lines, where the generators $J_A$ and $J_B$ are inserted at different positions).  We consider the
different irreducible pieces in turn.

\goodbreak
\noindent
\underline{\em $s$-type fluctuations}

\noindent
To study the irreducible piece with spin $\ell + 1$ we take $x$ to be symmetric and traceless on all indices,
\be
x_{AA_1 \cdots A_\ell} = s_{AA_1 \cdots A_\ell}
\ee
For such a tensor the Gauss law is automatically satisfied, while the equation of motion (\ref{xAeom3}) reduces to
\be
\ddot{s}_{AA_1 \cdots A_\ell} + {4 \over N^2} U^2 \ell(\ell - 1) s_{AA_1 \cdots A_\ell} = 0
\ee
We read off the frequencies
\be
\omega_\ell = {2 \over N} U \sqrt{\ell(\ell - 1)}
\ee
These modes are $(2 \ell + 3)$-fold degenerate.  There are two zero-frequency modes: $\ell = 0$ is a translation zero mode in the $X^A$ directions, while $\ell = 1$ is an
energy-preserving quadrupole deformation of the sphere.

\goodbreak
\noindent
\underline{\em $t$-type fluctuations}

\noindent
These exist for $\ell \geq 1$.
We can reconstruct the tensor $x$ from its spin-$\ell$ irreducible piece $t$ by setting
\be
\label{InverseMap}
x_{AA_1 \cdots A_\ell} = \epsilon_{AA_1B} t_{BA_2 \cdots A_\ell} + \epsilon_{AA_2B} t_{A_1 B A_3 \cdots A_\ell} + \cdots
+ \epsilon_{AA_\ell B} t_{A_1 \cdots A_{\ell - 1} B}
\ee
This map has been constructed so that $x$ is symmetric and traceless on the indices $A_1 \cdots A_\ell$.  In other words, it defines the embedding of
$(\hbox{\rm spin $\ell$}) \hookrightarrow (\hbox{\rm spin 1}) \otimes (\hbox{\rm spin $\ell$})$.  Given (\ref{InverseMap}), the corresponding Hermitian
matrix $x^A$ can be written as a commutator,
\bea
x^A & \equiv & x_{AA_1 \cdots A_\ell}  J_{A_1} \cdots J_{A_\ell} \\
\nonumber
& = & i  t_{A_1 \cdots A_\ell} [J_A, J_{A_1} \cdots J_{A_\ell}]
\eea
As we saw earlier, these fluctuations fail to satisfy the Gauss constraint, since from (\ref{FuzzyCasimir})
\be
[J^A,x^A] = i \ell (\ell + 1) t_{A_1 \cdots A_\ell} J_{A_1} \cdots J_{A_\ell}
\ee
Again the only solution to the Gauss constraint is to set $t = 0$.

\goodbreak
\noindent
\underline{\em $u$-type fluctuations}

\noindent
These exist for $\ell \geq 1$.  We can reconstruct $x$ from its spin-$(\ell - 1)$ irreducible piece using
\be
x_{AA_1 \cdots A_\ell} = \sum_{i = 1}^\ell \delta_{A A_i} u_{A_1 \cdots \widehat{A}_i \cdots A_\ell} - {2 \over 2 \ell - 1}
\sum_{\scriptstyle i,j = 1 \atop \scriptstyle i < j}^\ell \delta_{A_i A_j} u_{AA_1 \cdots \widehat{A}_i \cdots \widehat{A}_j \cdots A_\ell}
\ee
This map is constructed so that $x$ is symmetric and traceless on $A_1 \cdots A_\ell$.  For such a tensor the Gauss law is automatically
satisfied.  Substituting the expression for $x$ into the equation of motion
(\ref{xAeom3}), we find after some algebra that
\be
\ddot{u}_{A_2 \cdots A_\ell} + {4 \over N^2} U^2 (\ell+1)(\ell +2) u_{A_2 \cdots A_\ell} = 0
\ee
From this we read off the frequencies
\be
\omega_\ell = {2 \over N} U \sqrt{(\ell+1)(\ell+2)}
\ee
These modes are $(2\ell-1)$-fold degenerate.  The $\ell = 1$ mode is a monopole deformation of the sphere, $U \rightarrow U + \delta U$.  The frequency
$\omega_1$ agrees with what one obtains by perturbing the background equation of motion (\ref{BackgroundEOM}).

\section{Open string production\label{appendix:open}}

In this appendix we study the process of open string production in more detail.  Our goal is to show that, during the initial
collapse of a fuzzy sphere, roughly one open string is produced in each of the fluctuation modes.  We assume the fluctuations
are weakly coupled, which as discussed in \S \ref{sect:parametric} means $U_0 > N^{1/2} \lambda^{1/3}$.

We focus on a particular fluctuation mode.  For concreteness we consider a transverse mode (\ref{omega_ell}) with frequency
\be
\omega_\ell = {2 \over N} \sqrt{\ell(\ell+1)} \, U
\ee
For this mode, the adiabatic approximation breaks down when $\dot{\omega}_\ell / \omega_\ell^2 \sim 1$ or
\be
{N \dot{U} \over 2 \sqrt{\ell(\ell + 1)} \, U^2} \sim 1
\ee
Energy conservation (\ref{velocity}) fixes $\dot{U}^2 \approx {4 \over N^2} \big( U_0^4 - U^4 \big)$.  By the time the adiabatic
approximation has broken down we can neglect the $U^4$ term, so the velocity is
\be
\label{velocity2}
\dot{U} \approx {2 \over N} U_0^2
\ee
and the adiabatic approximation fails at
\be
\label{breakdown}
U \approx {U_0 \over \big(\ell(\ell+1)\big)^{1/4}}
\ee
At the point where the adiabatic approximation fails the mode can be thought of as a harmonic oscillator in its ground state, with a frequency
\be
\omega \approx {2 \over N} \big(\ell(\ell+1)\big)^{1/4} U_0
\ee
and a ground state wavefunction (identifying $\hbar / m$ with $g^2_{\rm YM}$)
\be
\label{psi_0}
\psi_0(x) = \left({\omega \over \pi g^2_{\rm YM}}\right)^{1/4} e^{-{1 \over 2} \omega x^2 / g^2_{\rm YM}}
\ee

After the adiabatic approximation breaks down the sphere continues to shrink.  We must follow the evolution of the mode through the
non-adiabatic regime, until the sphere re-expands to the radius (\ref{breakdown}) at which adiabaticity is restored.  In the non-adiabatic regime the
frequency is so low that it seems reasonable to neglect the potential energy for the mode, in other words, to treat it as a free particle.
In this approximation the Gaussian wavefunction (\ref{psi_0}) undergoes free diffusion, spreading to a width
\be
\Delta x^2 = \Delta x_0^2 + {g^4_{\rm YM} \Delta t^2 \over 4 \Delta x_0^2}
\ee
Here the initial position uncertainty is $\Delta x_0^2 = g^2_{\rm YM} / 2 \omega$, while the time spent in the non-adiabatic regime is
\be
\Delta t = {\Delta U \over \dot{U}} \approx {N \over \big(\ell(\ell+1)\big)^{1/4} U_0}
\ee
This gives $\Delta x^2 \approx 5 \Delta x_0^2$: the wavefunction spreads by a factor of roughly $\sqrt{5}$ as the sphere transits
the non-adiabatic regime.  This factor is independent of the parameters $N$, $\ell$, $U_0$, which suggests that of order one open string
is produced in each of the fluctuation modes.

To argue this more precisely we recall some properties of squeezed states \cite{optics}.  For a harmonic oscillator these are defined by
\be
\label{rep1}
\vert \xi \rangle = \exp \bigg[{{\xi \over 2} \left(\hat{a}^\dagger \hat{a}^\dagger - \hat{a} \hat{a}\right)}\bigg] \vert 0 \rangle
\ee
where the squeezing parameter $0 < \xi < \infty$.  An equivalent expression is
\be
\label{rep2}
\vert \xi \rangle = \left(1 - \gamma^2\right)^{1/4} \exp \bigg[{{\gamma \over 2} \hat{a}^\dagger \hat{a}^\dagger}\bigg] \vert 0 \rangle
\ee
where $\gamma = \tanh \xi$.  A squeezed state has a Gaussian wavefunction with a width
\be
\Delta x = e^\xi \Delta x_0
\ee
so we identify $e^\xi \approx \sqrt{5}$.  Expanding the exponential in (\ref{rep2}), the probability of finding $2n$ strings present is
\be
P(\hbox{\rm $2n$ strings}) = \left(1 - \gamma^2\right)^{1/2} {(2n)! \over (n!)^2} \, \left({\gamma \over 2}\right)^{2n}
\ee
The probability decreases monotonically with $n$.  The average number of strings present is
\be
\sum_{n = 0}^\infty 2 n P(\hbox{\rm $2n$ strings}) = {\gamma^2 \over 1 - \gamma^2} \approx {4 \over 5}
\ee
So the simple approximation of free diffusion in the non-adiabatic regime supports the claim that roughly one open string is produced in each
fluctuation mode.

\section{More on $U(t)$ oscillations\label{appendix:moreonU(t)}}

During the initial collapse, after $U$ has passed by $U = U_{\rm inner}$, we know that ${\cal O}(N^2)$ open strings have been created.
Then the dynamics of the fuzzy sphere radius $U(t)$ is dominated by 
the following energy conservation law:
\bea
&& \frac{2 U_0^4}{\lambda} = \frac{N^2}{2 \lambda} \dot{U}^2 + V_{\rm pot}(U)  \,, \nn \\
&& V_{\rm pot}(U) \equiv \frac{2 U^4}{\lambda} + c N^2 \vert U \vert \,.
\eea
Here $U_0$ sets the total energy. This is a one-dimensional oscillator
with a potential $V_{\rm pot}(U)$ which is positive definite and 
monotonically increasing as we increase $U$. It has linear behavior 
for small $U$, $U < U_c$ and $U^4$ behavior for large $U$, $U > U_c$ where  
\bea
\label{Ucdef}
U_c \equiv \left(\frac{c}{2}\right)^{1/3} N^{2/3}  \lambda^{1/3}\,.
\eea
In this appendix we study the resulting dynamics for $U(t)$ in more detail.
We will always consider $U_0$ satisfying 
$U_0 > N^{1/2} \lambda^{1/3}$
so that a perturbative analysis is valid. 

\subsection{Intermediate initial radius, $U_0 < U_c$}

In this case, the dynamics of $U$ is restricted to the region where the potential has linear behavior.
Keeping just the linear term in $V_{\rm pot}(U)$, the conservation law reads
\bea
\label{inter_conserv_law}
\frac{2U_0^4}{\lambda} 
&\approx& \frac{N^2}{2\lambda} \dot{U}^2 +  c N^2 \vert U \vert \,.
\eea
This sets the amplitude of oscillation of $U(t)$ as 
\be
\widetilde{U}_0 \approx \frac{2 U_0^4}{c \lambda N^2} 
\ee
Since $U_0 < U_c$, this yields a consistent relation
\be
\widetilde{U}_0  \sim \frac{U_0^4}{\lambda N^2}  < \frac{U^4_c}{\lambda N^2} \sim U_c \,.
\ee

Using $\widetilde{U}_0$, we can rewrite the conservation law (\ref{inter_conserv_law}) as 
\be
 \frac{N^2}{2 \lambda} \dot{U}^2 \approx  c N^2 ( \widetilde{U}_0 - U )  \,.
\ee
This yields periodic oscillations with period $\tau$, where
\be
\tau \approx  \sqrt{\frac{{ 32 \widetilde{U}_0}}{ c \lambda}}
\ee
or, neglecting some numerical factors including $c$, $\tau \sim \frac{U_0^2}{\lambda N}$.

\begin{figure}[t]
\begin{center}
\includegraphics[width=8cm]{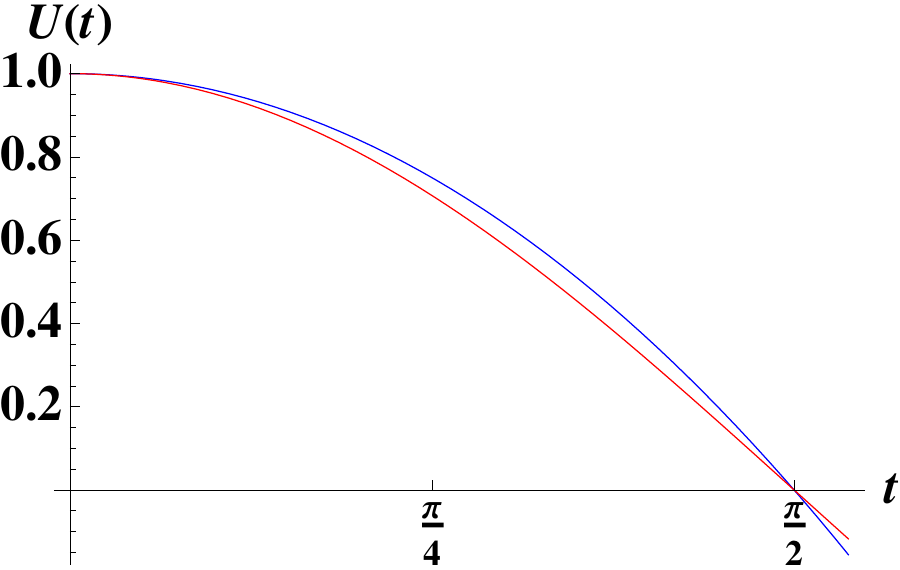}
\end{center}
\caption{Blue function: $U(t) = - (2 t/\pi)^2 + 1$, red function: $U(t) = \cos t$. The two functions are very similar.}
\label{t2vscos}
\end{figure}

Note that $U$ has acceleration or deceleration $c \lambda$, with periodicity $4 \sqrt{\frac{{ 2 \widetilde{U}_0}}{ c \lambda}}$, so
setting $t = 0$ at $U = 0$ we have
\[
U(t) = \left\lbrace
\begin{array}{ll}
\widetilde{U}_0  - \frac{1}{2} c \lambda \Big(t - (4 n + 1) \sqrt{\frac{{ 2 \widetilde{U}_0}}{ c \lambda}} \Big) ^2 \quad & \mbox{for} \quad
4n  \sqrt{\frac{{ 2 \widetilde{U}_0}}{ c \lambda}}  \le t \le (4 n + 2)  \sqrt{\frac{{ 2 \widetilde{U}_0}}{ c \lambda}} \\[6pt]
- \widetilde{U}_0 +  \frac{1}{2} c \lambda \Big(t - (4 n +3) \sqrt{\frac{{ 2 \widetilde{U}_0}}{ c \lambda}} \Big)^2 
\quad  & \mbox{for} \quad (4 n + 2)  \sqrt{\frac{{ 2 \widetilde{U}_0}}{ c \lambda}} \le t \le  (4 n + 4) \sqrt{\frac{{ 2 \widetilde{U}_0}}{ c \lambda}}
\end{array}\right.
\]
for integers $n$. This periodic behavior of $U(t)$ is well-approximated by a circular function with frequency $\Omega$ 
\be
\label{approxsinforUintermediate}
U(t) \approx \widetilde{U}_0 \sin \Omega t \,, \quad \widetilde{U}_0 \sim \frac{U_0^4}{\lambda N^2} \,, \quad
\Omega \sim \tau^{-1} \sim \frac{\lambda N}{U_0^2} \,. 
\ee
Figure \ref{t2vscos} shows that this approximation works very well. 

\subsection{Large initial radius, $U_0 > U_c$}

In this case, the dynamics of $U$ is no longer restricted to the region where the potential $V_{\rm pot}(U)$ 
has linear behavior. Instead the conservation law gives
\be
\frac{2 U_0^4}{\lambda} = \frac{N^2}{2 \lambda} \dot{U}^2 +  \frac{2 U^4}{\lambda}  + c N^2 U  \,.
\ee
Again we have oscillatory behavior.  Near the turning point
\be
\frac{U_0^4}{\lambda}  > N^2 U_0 
\ee
so the amplitude is very well approximated by $U_0$, $\widetilde{U}_0 \approx U_0$.

During each oscillation $U$ starts from $U_0$, passes by $U = U_c$ given in Eq.~(\ref{Ucdef}), 
then enters the region where $U < U_c$. 
The timescale for $U$ to run from $U= U_0$ to $U=U_c$, which we call   
$\Delta \tau_1$, is given by
\be
\Delta \tau_1 = 
N \int^{U_0}_{U_c} \frac{dU}{\sqrt{U_0^4 - U^4 - \lambda N^2 U}}
\sim N \int^{U_0}_{U_c} \frac{dU}{\sqrt{U_0^4 - U^4 }} \sim \frac{N}{U_0} \,, 
\ee
since in this region the potential $V_{\rm pot}(U)$ is well approximated by the quartic term.
On the other hand, the timescale for $U$ to run from $U= U_c$ to $U=0$, which we call 
$\Delta \tau_2$, is
\be
\Delta \tau_2 = N \int^{U_c}_{0} \frac{dU}{\sqrt{U_0^4 - U^4 - \lambda N^2 U}}  
\lesssim N \int^{U_c}_{0} \frac{dU}{\sqrt{U_0^4 }} = N \frac{U_c}{U_0^2} \,.
\ee
since in this region the potential is well approximated by the linear term.

\begin{figure}[t]
\begin{center}
\includegraphics[width=9cm]{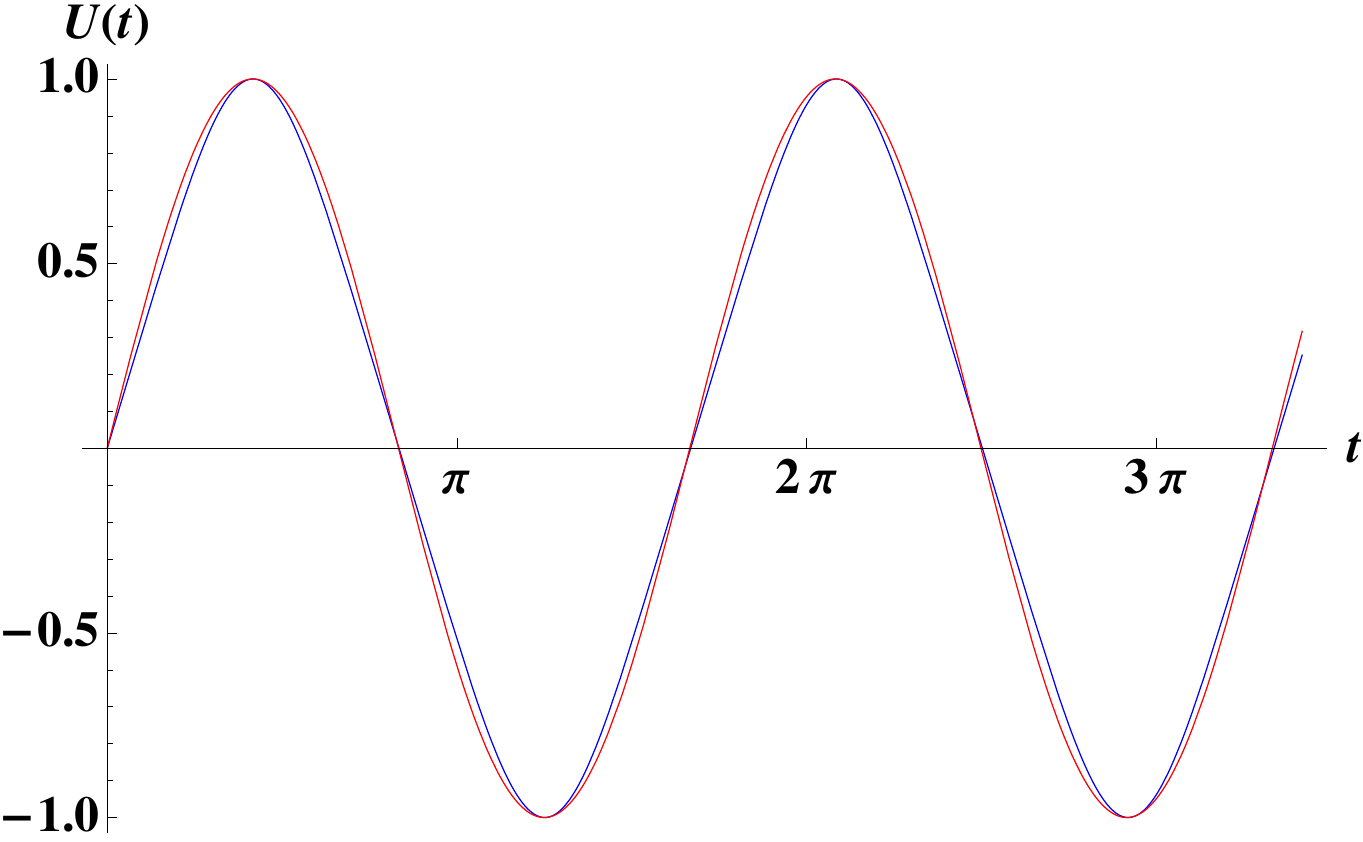}
\end{center}
\caption{Blue function: $U(t) = \text{sn}(t,-1)$, red function: $U(t) = \sin (1.2 t)$. The two functions are very similar.}
\label{JacobiSN}
\end{figure}

Since $U_c < U_0$, note that $\Delta \tau_1 > \Delta \tau_2$, and
therefore the period of oscillation is dominated by the motion from $U_0$ to $U_c$.
The conservation law is well approximated by neglecting the back reaction from open string creation and taking
\be
\frac{2 U_0^4}{\lambda} = \frac{N^2}{2 \lambda} \dot{U}^2 + \frac{2 U^4}{\lambda} \,.
\ee
Taking $t = 0$ at $U = 0$ we find the solution
\be
U(t) = U_0 \, \text{sn}\left(\frac{2 U_0 t}{N}, -1\right)
\ee
where $\text{sn}(u,m)$ is a Jacobi elliptic function, given by
\be
\text{sn}(u,m) = \sin \phi \quad \mbox{where $\phi$ is defined by} \quad u = \int_0^\phi \frac{ds}{\sqrt{1 - m \sin^2 s}} \,  \,.
\ee
In our case $m=-1$ so $u = \int_0^\phi \frac{ds}{\sqrt{1 + \sin^2 s}}$. 
Since $\text{sn}(u,m) = \sin \phi$ is periodic under $\phi \sim \phi + 2 \pi$, it follows that $u$ is periodic under
$u \sim u + \int_0^{2 \pi} \frac{ds}{\sqrt{1 + \sin^2 s}} \approx u + 2 \pi/1.2 $. In fact, the behavior of $U(t)$ is very well approximated by
\be
U(t) = U_0 \, \text{sn}\left(\frac{2 U_0 t}{N},-1\right) \approx U_0 \sin {\frac{2.4 U_0 t}{ N}} \,.
\ee
Figure \ref{JacobiSN} shows that this approximation works very well. This means $U(t)$ can be approximated as 
\be
\label{approxsinforUlarge}
U(t) \approx \widetilde{U}_0  \sin \Omega t \,, \quad 
\widetilde{U}_0 \sim U_0 \,, \quad
\Omega \sim  \tau^{-1} \sim \frac{U_0}{N} \,. 
\ee
Note that (\ref{approxsinforUintermediate}) and (\ref{approxsinforUlarge}) agree at  $U_0 \sim U_c$. 

\providecommand{\href}[2]{#2}\begingroup\raggedright\endgroup

\end{document}